\documentstyle[aps,twocolumn,psfig]{revtex}
\draft
\preprint{LA-UR-98-4014}

\newcommand{\lc}       {lanthanum cuprate}

\newcommand{\lsco}     {$\rm La_{2-x} Sr_x CuO_4$}
\newcommand{\lbco}     {La$_{2-x}$Ba$_x$CuO$_4$}
\newcommand{\lesco}    {La$_{1.8-x}$Eu$_{0.2}$Sr$_x$CuO$_4$}
\newcommand{\sample}   {La$_{1.67}$Eu$_{0.2}$Sr$_{0.13}$CuO$_4$}
\newcommand{\lnsco}    {La$_{2-x-y}$Nd$_y$Sr$_x$CuO$_4$}
\newcommand{\lescoy}   {La$_{2-x-y}$Eu$_y$Sr$_x$CuO$_4$}
\newcommand{\lmsco}    {La$_{2-x-y}$M$_y$Sr$_x$CuO$_4$}
\newcommand{\cuot}     {CuO$_2$}

\newcommand{\tlt}      {$T_{\rm LT}$}

\newcommand{\la}       {$^{139}$La}
\newcommand{\laslr}    {$^{139}T_1^{-1}$}
\newcommand{\cu}       {$^{63}$Cu}
\newcommand{\cuslr}    {$^{63}T_1^{-1}$}
\newcommand{\curt}     {$(^{63}T_1T)^{- 1}$}
\newcommand{\ssr}      {$^{63}T_{2G}^{-1}$}
\newcommand{\ltot}     {LTO $\rightarrow$ LTT}

\title{Spin Dynamics in the LTT Phase of ${\mathbf \cong 1/8}$ Doped Single
Crystal La$_{1.67}$Eu$_{0.2}$Sr$_{0.13}$CuO$_4$}

\author{B. J. Suh,$^{1,2}$ P. C. Hammel,$^1$ M. H\"{u}cker,$^3$ B.
B\"{u}chner,$^3$ U. Ammerahl,$^{3,4}$ and A. Revcolevschi$^4$}

 \address{$^1$Condensed Matter and Thermal Physics,
 Los Alamos National Laboratory, Los Alamos, NM 87545}
 \address{$^2$Catholic University of Korea, South Korea.}
 \address{$^3$II. Physikalisches Institut, Universit\"{a}t zu K\"{o}ln, 50937 K\"{o}ln, Germany}
 \address{$^4$Laboratoire de Chimie des Solides, Universit\'{e} Paris-Sud, 91405 Orsay Cedex, France\vspace{-4mm}}

\author{\small(Received:  \hspace{1cm}\quad)}

\address{\parbox{14cm}{\bigskip\rm\small We present \la\ and \cu\
NMR relaxation measurements in single crystal
La$_{1.67}$Eu$_{0.2}$Sr$_{0.13}$CuO$_4$. A strong peak in the \la\
spin-lattice relaxation rate observed in the spin ordered state is
well described by the BPP mechanism\protect\cite{bpp} and arises
from continuous slowing of electronic spin fluctuations with
decreasing temperature; these spin fluctuations exhibit $XY$-like
anisotropy in the ordered state. The spin pseudogap is
significantly enhanced by the static charge-stripe order in the
LTT phase.
 \\ 1998 PACS numbers:  76.60.Jx, 74.25.Ha, 76.60.-k, 74.72.Dn }}

\begin{document}
\maketitle

\thispagestyle{myheadings} \markright{{\em LA-UR-98-4014}}

Understanding spin dynamics and correlations in high temperature
superconductors (HTSC) is crucial to solving the mechanism of the
superconductivity. The existence of low-frequency
antiferromagnetic (AF) spin fluctuations in the \cuot\ planes and
the opening of spin pseudogap in the normal state are two poorly
understood features of HTSC\cite{brinkmann,oki}. An anomalous
suppression of $T_c$ is observed for a hole concentration of 1/8
in \lbco\ and for a range of hole concentrations near this value
in rare-earth co-doped \lmsco\ (M = Eu, Nd)\cite{axe:prl,bu:prl};
in both cases a structural phase transition (SPT) to the low
temperature tetragonal (LTT) phase occurs. The observation of
incommensurate elastic neutron diffraction peaks indicative of
static charge stripe order in the LTT phase of \lnsco \cite{tr}
followed by incommensurate magnetic order indicates that spin
order is induced by the charge-stripe order\cite{tr} and
emphasizes the importance of understanding the magnetism in these
charge-stripe ordered systems.

Incommensurate magnetic peaks have also been observed near $\sim
30$ K in Eu co-doped compounds by elastic neutron
scattering\cite{NSlesco} revealing static stripe order, and muon
spin rotation ($\mu$SR) studies confirm magnetic order in Nd and
Eu compounds below $\sim 30$ K\cite{nd:usr,nachumi:ndusr,eu:usr}.
$\mu$SR studies of the Nd material find a lower magnetic ordering
temperature indicating quasistatic magnetic behavior, and
Tranquada {\it et al.} have argued that the magnetic order is
glassy\cite{tr:glassy}. Electron spin resonance (ESR) studies in a
series of \lescoy\ samples\cite{krb} have associated this ordering
with continuous slowing of spin fluctuations with decreasing
temperature.

\begin{figure}[tb]
   \parbox{3.3in}{
   \psfig{file=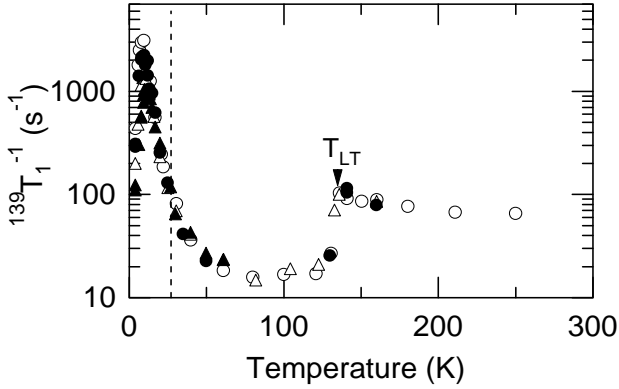,width=3.2in}}
   \vspace{2mm}
   \caption{The lanthanum spin-lattice relaxation rate \laslr\ in single
   crystal \sample\ measured in two applied fields ($\nu$ = 29 and 50 MHz)
   is shown for two orientations of the field ({\bf H}) with respect to the
   crystal $c$-axis.}
   \label{fig:lat1}
   \end{figure}

We have investigated the spin dynamics of $x \cong 1/8$ doped
\lesco\ using nuclear magnetic resonance (NMR) to better
understand the influence of charge-stripes on the quasistatic spin
order. We report a strong peak in the \la\ spin-lattice relaxation
rate \laslr\ below the spin ordering temperature (see
Fig.~\ref{fig:lat1}), a remarkable situation quite similar to AF
ordered \lc\ with an order of magnitude smaller
doping\cite{chou,borsa,suh:li}. The frequency dependence of this
peak is well explained by a mechanism first discussed by
Bloembergen, Purcell and Pound (BPP): \la\ relaxation is due
electronic spin fluctuations whose characteristic rate
$\tau_c^{-1}$ slows with decreasing temperature, and the peak
occurs (at $T=T_f \sim 9$ K) when $\tau_c^{-1}$ matches the NMR
frequency (30--50 MHz). Our measurements also reveal a very strong
orientation dependence of \laslr\ in the vicinity of the peak
demonstrating a strong anisotropy of the spin fluctuations in the
ordered state similar to those observed in $XY$-like systems.
\laslr\ also exhibits an abrupt decrease at \tlt. \cu\
spin-lattice and spin-spin relaxation measurements reveal a spin
pseudogap significantly more pronounced than occurs in similarly
doped \lsco\ indicating that static charge stripes in the LTT
phase significantly enhance the spin pseudogap.

Our \la\ and \cu\ NMR relaxation measurements were performed on a
La$_{1.67}$Eu$_{0.2}$Sr$_{0.13}$CuO$_4$ single crystal which
undergoes the \ltot\ SPT at $T = 134 \pm 2$~K. The crystal was
grown using the traveling solvent floating zone method under
oxygen pressure of 3 bar\cite{ammer}. From dc magnetization
measurements, no superconducting transition is observed down to
4.2~K. Static spin order is observed below $\approx 20$~K in
$\mu$SR studies\cite{eu:usr}. Both the \la\ ($I= {7 \over 2}$) and
\cu\ ($I = {3 \over 2}$) spin-lattice relaxation rates were
measured by monitoring the recovery of the central transition
(\(m_I = + {1 \over 2} \leftrightarrow - {1 \over 2} \))
magnetization after saturation with a single ${\pi/ 2}$ pulse. The
\cu\ spin-spin relaxation was measured by monitoring the spin-echo
decay using a two-pulse Hahn-echo (\(\frac{\pi} {2} \mbox{-} \tau
\mbox{-} \pi\)). The time dependence of the magnetization recovery
does not conform to the standard theoretical
expression\cite{recovery} over the entire temperature range
investigated. Far above \tlt\ and in the intermediate range $30
\lesssim T \lesssim \,$\tlt, the data were well fit by this
expression for magnetic relaxation and saturation by a single
pulse\cite{recovery}, however around \tlt\ and at low $T$ ($
\lesssim 30$~K), this fit is poorer. We ascribe the slight
deviation around \tlt\ to the additional motion of oxygen
octahedra associated with the SPT; this generates a quadrupolar
contribution to the \la\ relaxation\cite{suh:eunqr}. As in lightly
doped La214\cite{chou,suh:li,suh:eunqr}, stretched exponential
recoveries are observed at low $T$.
   \noindent
To provide a consistent basis for analysis of the $T$-dependence
of \laslr\, the first decade of recovery data was fit, for all
$T$, to the stretched exponential function
\([M(\infty)-M(t)]/M(\infty) = \exp[-(t/T_1)^{1/2}]\). While this
analysis increases the uncertainty in \laslr\ for $T \gtrsim
30$~K, we find that varying the fitting procedure has essentially
no effect on the behavior at \tlt.

\begin{figure}[tb]
   \parbox{3.4in}{
   \psfig{file=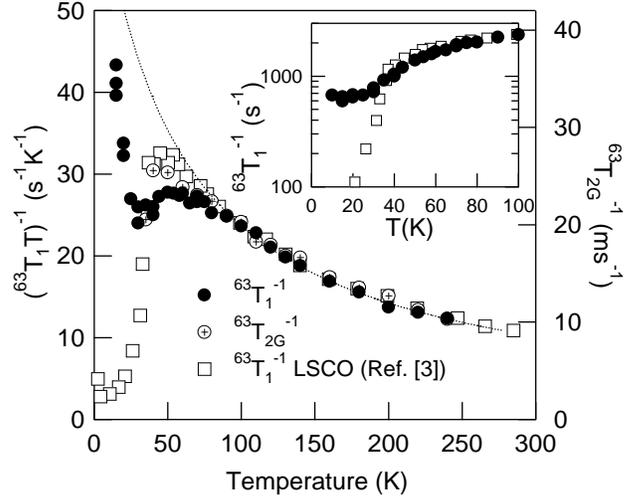,width=3.2in}}
   \vspace{2mm}
   \caption{Copper spin-lattice relaxation: the $T$-dependencies of \curt\
   and \ssr\ measured in an applied field ($\nu = 90$~MHz) are shown for
   ${\bf H} \parallel c$-axis; \curt\ for identically doped $\rm La_{1.87}
   Sr_{0.13} Cu O_4$ (Ref.~\protect\onlinecite{oki}) is also shown for
   comparison. The solid curve is a Curie-Weiss law fit for $T \geq 100$~K:
   \curt\ $\propto 1/(T+\theta)$ with $\theta = 16$~K. Inset: A semilog
   plot of \cuslr\ vs.\ $T$.}
   \label{fig:cut1}
   \end{figure}

The $T$-dependencies of \curt\ and \ssr\ are shown in
Fig.~\ref{fig:cut1}. The \cu\ magnetization recovery data for
whole temperature range investigated are well fit by the
theoretical expression\cite{recovery} for magnetic relaxation
following single pulse saturation: $[M(\infty)-M(t)]/M(\infty) =
0.1 \exp(-t/T_1) + 0.9 \exp (-6t/T_1)$. \ssr\ was obtained by
fitting the spin-echo amplitude, $S(t\mbox{=}2\tau)$, to the
expression: $S(t)=S(0)\exp (-t/T_{2R}) \exp [-(t/T_{2G})^2/2]$.
The contribution of spin-lattice relaxation processes to the spin
echo decay, $T_{2R}^{-1}$, was determined from $T_{2R}^{-1} =
(\beta+R)/^{63}T_1$ with $\beta = 3$ and the anisotropy of \cuslr
, $R = 3.6$\cite{imai:tg}. Because we cannot entirely invert the
Cu line these $T_{2R}^{-1}$ data are not quantitatively accurate,
but they reliably indicate the qualitative $T$-dependence of
$T_{2R}^{-1}$.
The \cu\ NMR linewidth (full width at half maximum) $\Delta H
\simeq 1.2$~kG at $T = 240$~K, and is observed to increase
monotonically to $\simeq 3.4$~kG at $T = \,\,$40 K. Consistent
with earlier work we observe a strong suppression of the intensity
of the \cu\ NMR signal with decreasing $T$\cite{hunt}; this
suppression is well explained by the slow electron spin
fluctuations responsible for the low $T$ peak in \laslr
\cite{curro:lesco}.

In our discussion we will focus on the following: (i)
characterizing the \ltot\ SPT, (ii) demonstrating that the low $T$
spin-freezing peak in \laslr\ is well described by the BPP
mechanism\cite{bpp} which reveals that the low-$T$ spin dynamics
are characterized by a {\it distribution} activation energies,
$E_a/k_B \sim 100$ K , and (iii) \cuslr; particularly the contrast
with $x=1/8$, LTO-phase superconducting \lsco.

The abrupt decrease in \laslr\ at \tlt\ with essentially no
enhancement above \tlt\ contrasts with the behavior
   \noindent
of lightly doped \lesco, where a strong enhancement is observed
above \tlt \cite{suh:eunqr}. This result is consistent with the
observation that the \ltot\ SPT in this heavily doped sample is
first order\cite{bu:euro,axe:jltp}.

\begin{figure}[t]
   \parbox{3.3in}{
   \psfig{file=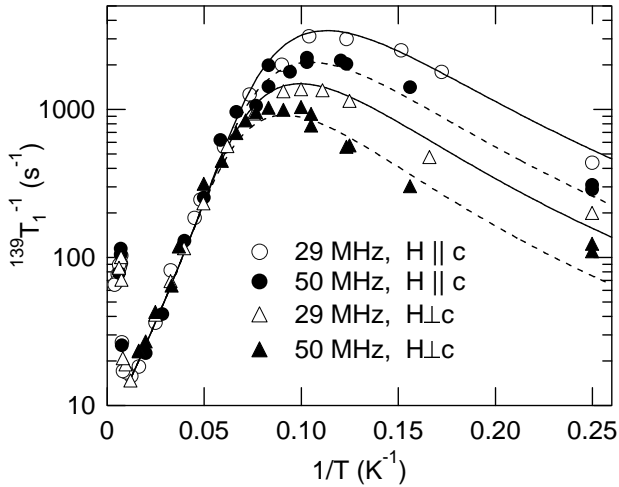,width=3.2in}}
   \vspace{2mm}
   \caption{\laslr\ vs.\ $1/T$.  The curves are theoretical fits as
   described in the text using the fitting parameters: $E_0/k_B =
   82$~K, $\Delta = 25$~K for both orientations and frequencies; for
   \(H \! \parallel \! c {\text :} \,\, \tau_{\infty} = 1.4 \times
   10^{-13}\) s and \(C = 1.67 \times 10^{12}\, {\rm s}^{-2}\), and
   for \(H \! \perp \! c {\text :} \,\, \tau_{\infty} = 3.1 \times
   10^{-13}\)~s and \(C = 0.73 \times 10^{12}\, {\rm s}^{-2}\).}
   \label{fig:bpp}
   \end{figure}

The strong orientation and frequency dependencies of the low $T$
peak in \laslr\ evident in Fig.~\ref{fig:bpp} clearly indicate
that it results from continuous slowing of anisotropic spin
fluctuations with decreasing $T$. The same frequency dependence of
\laslr\ below $T_f$ is observed by comparing different NQR
transitions in zero field, hence the frequency dependence does not
reflect a magnetic field dependence. The nuclear spin-lattice
relaxation rate is given by \( T_1^{-1} \propto \gamma ^2 h_{\rm
0}^2 \jmath(\omega)\)\cite{CPS} where $\jmath$ is the spectral
density function of spin fluctuations, $\omega = 2\pi\nu$ is the
NMR frequency, and $h_{\rm 0}$ is the fluctuating component of the
effective hyperfine field perpendicular to the applied field.

As we show in Fig.~\ref{fig:bpp}, the magnitude of \laslr\ also
depends strongly on the field orientation in the vicinity of
$T_f$: \( [^{139}T_{1}^{-1}(H \parallel c)] / [^{139}T_1^{-1}(H
\perp c)] \cong 2.3 \pm 0.3\) (see the ratio of $C$ in
Fig.~\ref{fig:bpp}), near 2, the value expected for $XY$-like spin
fluctuations, where the fluctuations of the out-of-plane component
are entirely frozen\cite{suh:cl}. This striking $XY$-like behavior
of spin fluctuations observed in this heavily doped system is
reminiscent of the crossover from Heisenberg to $XY$-like observed
at temperatures just above the onset of long range
antiferromagnetic order in \(\rm Sr_2CuO_2Cl_2\)\cite{suh:cl}.

The frequency and temperature dependencies of \laslr\ are well
explained by the well-known BPP mechanism\cite{bpp} that describes
the effects of continuous slowing on the spin-lattice relaxation,
although the anisotropic fluctuations and asymmetric peak evident
in Fig.~\ref{fig:bpp} are beyond the standard BPP picture. In
particular, the
   \noindent
stretched exponential recovery of the magnetization and the
asymmetric peak implies a distribution of characteristic
correlation times ($\tau_c$) of the fluctuations\cite{fastion}. We
take $\tau_c(E_a,T) = \tau_{\infty} \exp (E_a/k_BT)$\cite{bpp},
however we must include a distribution of activation energies
$E_a$ (a Gaussian works well): \(Z(E_a) = (\sqrt{ 2
\pi}\Delta)^{-1} \exp \left[- (E_a-E_0)^2/2(k_B \Delta)^2 \right
]\).   For two-dimensional diffusive fluctuations (in which
$c$-axis fluctuations are frozen), we can write\cite{suh:hg}:
    \(\jmath(\omega) = \tau_c \ln \left [(\tau_c^{-2} + \omega^2)/ \omega^2 \right]\).
However the much more general Lorentzian spectral density
(applicable to fluctuations of all wavelengths) provides equally
good fits with small changes in parameter values ($E_0$ and
$\Delta$ differ by less than 20\%). The measured relaxation rate
will be an average over the distribution $Z(E_a)$:
  \begin{equation}
     {T_1^{-1}}(T)=C\int_{0}^{\infty}\tau_c\ln \left( \frac{\tau_c^{-2} + \omega^2}{\omega^2}\right) Z(E_a)dE_a \,,
  \end{equation}
where the coefficient $C$ is proportional to $h_{\rm 0}^2$. The
fits shown in Fig.~\ref{fig:bpp} demonstrate that the BPP picture
(with a distribution of $E_a$) accurately describes \laslr at low
$T$, accounting well for the frequency dependence evident at
temperatures below the peak.  This provides direct evidence for
the {\it intrinsic role of disorder in the continuous slowing of
the spin fluctuations.} Surprisingly, we find the ratio
\([\tau_{\infty}^{-1}(H \! \parallel \! c)] /
[\tau_{\infty}^{-1}(H \! \perp \! c)] \cong 2.2 \) is essentially
equal to the ratio of \laslr\ (and the ratio of $C$). This is
apparently related to the intrinsic anisotropy of two dimensional
spin system\cite{suh:cl}, although it is not understood
theoretically at present. (We note that studies of the
stripe-ordered state of $\rm La_{5/3}Sr_{1/3}NiO_4$ revealed no
dependence of the static ordered magnetic moment on the
orientation and magnitude of the applied magnetic
field\cite{yoshinari:prl99}.) The value of $\tau_{\infty}$ cannot
be extracted from these fits with high precision (\(0.03 \mbox{
psec} \! < \! \tau_{\infty} \! < \! 0.3 \) psec), however the
uncertainty in $E_0$ is much smaller (\(82  \! < \! E_0/k_B \! <
\! 96\) K). These slow fluctuations are in qualitative agreement
with the slowing of fluctuations observed in ESR
measurements\cite{krb}.

The strong similarity of the spin-freezing observed in the present
metallic sample to that seen in lightly doped \lc\ is surprising
given the broad range of doping involved and the strong consequent
variation in magnetic properties. Scenarios have been proposed
attributing spin freezing and the associated recovery of the
sublattice magnetization observed in lightly hole-doped La214 to
freezing of domain motion\cite{suh:li,mobile} or to the effective
disappearance of domain boundaries as the constituent holes become
pinned to the lattice at low $T$\cite{borsa}. These do not appear
applicable here: (i)~Our heavily doped sample is a fair conductor
whose resistivity $\rho$ decreases with decreasing $T$ to fairly
low $T$ with only a weak increase and a broad maximum below
\tlt\cite{bu:pc}. The magnitude of $\rho$ itself is of order
$10^{-3} \, \Omega$-cm, one or two orders of magnitude smaller
than in lightly doped La214. (ii)~Neutron scattering shows the
domain boundaries remain in the spin-freezing regime\cite{tr}.

Two classes of mechanisms that could explain the spin freezing
data in the present heavily doped sample can be considered. The
first is related to scenarios previously discussed in the context
of lightly-doped \lc\cite{cho:nqr,neto,vdz}; the low temperature
freezing could reflect the behavior of an ordered antiferromagnet
spatially interrupted by an array of domain walls. The second
involves the dynamics of charged domain
walls\cite{kivelson:nature,zaanen,neto:prl99}. This motion will
alter the local magnetization in the spin ordered domains; the
spin freezing would result from the gradual suppression of the
excitations of this coupled system with decreasing temperature.
The doping (and hence stripe density) {\it in\/}dependence of the
freezing suggests that charge-stripe dynamics are not
responsible\cite{curro:lesco}. The distribution of activation
energies we observe indicates that stripe defects and disorder are
important.

We consider the first of these possibilities in the present more
heavily-doped sample: the spin freezing would reflect the slowing
of fluctuations that would result as spin domains separated by
charged domain walls become coupled allowing the correlation
length to grow. At high $T$ the coupling $J'$ between individual
spins separated by domain walls is likely weak compared to the
exchange coupling constant $J$ within a domain. However, the
occurrence of long range order at low $T$ with correlation lengths
long compared to the stripe spacing\cite{tr} indicates significant
effective coupling between neighboring AF domains. The strength of
the coupling between adjacent domains is proportional to $J'\xi^z$
where the spin correlation length $\xi$ is a function of $J$ and
the exponent $z$ is close to unity depending upon the
dimensionality of domains. The very slow characteristic timescale
(\(\sim 10^{-13} \) s) and the $XY$-like character of the spin
fluctuations we observe are consistent with a very large $\xi$
above the apparent spin ordering temperature ($\approx 20$~K). At
some low $T$, $J'\xi^z$ would become large enough ($\sim k_BT$) to
couple neighboring domains; the characteristic fluctuation rate
$\tau_c^{-1}$ will slow as the size of the correlated regions
grow.

We now turn to the behavior of \cuslr\ which reflects the AF spin
fluctuations and correlations. As shown in Fig.~\ref{fig:cut1},
\curt\ exhibits Curie-Weiss behavior at high temperature above the
opening of the spin pseudogap. The opening of the spin pseudogap
is evident as a reduction of \curt\ compared to the Curie-Weiss
behavior resulting in a broad peak in \curt. The difference in the
temperature dependencies of \curt\ and \ssr\ below the peak
position is indicative of the opening of a dynamic spin gap at
$q=q_{AF}$\cite{brinkmann,julien}. Fig.~\ref{fig:cut1} shows
\cuslr\ in \lc\ both with and without the Eu co-doping; the
behavior is essentially identical down to $\approx 50$~K), while
below this \lesco\ exhibits a substantially enhanced spin gap. It
has been argued that a spin gap arises naturally in undoped spin
ladders\cite{ekz,scalapino}; the enhanced spin gap observed in the
LTT phase may arise from the improved confinement of the undoped
domains into ladder-like structures by the static charge stripes
at commensurate doping. In the spin freezing region below $\approx
20$~K, \cuslr\ approaches a constant value in contrast to the
rapid decrease in the superconducting \lsco \cite{oki}
(Fig.~\ref{fig:cut1}, inset); consistent with the absence of a
superconducting transition in this spin freezing region.

We have presented a \la\ and \cu\ NMR study of the spin dynamics
in 1/8 doped, stripe-ordered \sample. We show that the strong peak
in \laslr\ occurring below the AF ordering temperature is due to
the well known BPP mechanism\cite{bpp}. The continuous slowing of
spin fluctuations is characterized by a distribution of activation
energies indicating the role of dynamical disorder in the
previously noted quasi-static behavior\cite{nachumi:ndusr}, in
addition to static disorder\cite{tr:glassy}. We find the opening
of the spin pseudogap is more pronounced in the presence of the
static charge stripes at commensurate doping in the LTT phase.
This suggests that the spin gap may be associated with the
confinement of the spin regions separated by hole-rich domain
walls.

We thank F. Borsa, A.H. Castro Neto, S.A. Kivelson and D.J. Scalapino for
helpful suggestions. The work at Los Alamos was performed under the
auspices of US Department of Energy. The work at University of K\"{o}ln
was supported by the Deutsche Forschungsgemeinschaft through SFB~341.
M.H. acknowledges support by the Graduiertenstipendium des Landes
Nordrhein-Westfalen.


\end{document}